\documentclass[]{aa}
\usepackage{graphicx,subfigure,natbib,url,txfonts}
\bibpunct{(}{)}{;}{a}{}{,}

\begin{document}
\title{Solar irradiance variability: A six-year comparison between SORCE observations and the SATIRE model}

\author{Will T. Ball\inst{1}, Yvonne C. Unruh\inst{1}, Natalie A. Krivova\inst{2}, Sami Solanki\inst{2,4}, Jerald W. Harder\inst{3}}
        \offprints{W T Ball}
	\institute{Astrophysics Group, Blackett Laboratory, Imperial College London, SW7 2AZ, United Kingdom
   \and 
	Max-Planck-Institut f\"ur Sonnensystemforschung, D-37191 Katlenburg-Lindau, Germany 
   \and
	Laboratory for Atmospheric and Space Physics, 1234 Innovation Drive, Boulder, 
	Colorado 80303-7814, USA
	\and
	School of Space Research, Kyung Hee University, Yongin, Gyeonggi, 446-701, Korea
              }
\date{Received \today; accepted }

\abstract
{}
{We investigate how well modeled solar irradiances agree with measurements from the SORCE satellite, both for total solar irradiance and broken down into spectral regions on timescales of several years.
}
{We use the SATIRE model and compare modeled total solar irradiance (TSI) with TSI measurements over the period 25 February 2003 to 1 November 2009. Spectral solar irradiance over 200-1630 nm is compared with the SIM instrument on SORCE over the period 21 April 2004 to 1 November 2009. We discuss the overall change in flux and the rotational and long-term trends during this period of decline from moderate activity to the recent solar minimum in $\sim$10 nm bands and for three spectral regions of significant interest: the UV integrated over 200-300nm, the visible over 400-691 nm and the IR between 972-1630 nm.
}
{The model captures 97\% of the observed TSI variation. This is on the order at which TSI detectors agree with each other during the period considered. In the spectral comparison, rotational variability is well reproduced, especially between 400 and 1200 nm. The magnitude of change in the long-term trends is many times larger in SIM at almost all wavelengths while trends in SIM oppose SATIRE in the visible between 500 and 700 nm and again between 1000 and 1200 nm. We discuss the remaining issues with both SIM data and the identified limits of the model, particularly with the way facular contributions are dealt with, the limit of flux identification in MDI magnetograms during solar minimum and the model atmospheres in the IR employed by SATIRE. However, it is unlikely that improvements in these areas will significantly enhance the agreement in the long-term trends. This disagreement implies that some mechanism other than surface magnetism is causing SSI variations, in particular between 2004 and 2006, if the SIM data are correct. Since SATIRE was able to reproduce UV irradiance between 1991 and 2002 from UARS, either the solar mechanism for SSI variation fundamentally changed around the peak of cycle 23, or there is an inconsistency between UARS and SORCE UV measurements. We favour the second explanation.
}
{}


\titlerunning{Comparison between SORCE observations and the SATIRE model}
\authorrunning{Ball et al.}
\maketitle

\section{Introduction} 
%
\label{sec:intro}

It has now been well established that over the last 30 years of recorded total solar irradiance (TSI) the Sun has varied by $\sim$0.1\% over an 11-year solar cycle. In the past it was assumed that the radiative forcing on Earth was directly proportional to this change in the total solar output, and although there is evidence that changes in TSI do affect the climate \citep{LabitzkevanLoon1995,vanLoonShea1999}, it is becoming clear that spectral solar irradiance (SSI) is also an important factor when considering the Sun's impact on climate \citep{Haigh1994}. It has previously been estimated that 60\% of TSI variability is found in ultra-violet (UV) wavelengths below 400 nm \citep{KrivovaSolanki2006}. The UV domain is known to affect stratospheric temperatures and chemistry; most notable and well-established is its effect on ozone \citep{LabitzkeAustin2001,Haigh2007}.

UV spectral observations extend back to 1978 providing a long-term record of variability over 120-400 nm. A composite has been produced by \cite{DeLandCebula2008}. However, this only accounts for absolute differences between different instruments and does not take into account instrumental problems and trends and so requires further adjustments before being truly useful for climate studies. 

Although the VIRGO/SPM instrument onboard the SoHO spacecraft has been observing three narrow spectral bands in the visible region since 1996, a wide band record of wavelengths longer than the UV has only recently started to accumulate with the SCIAMACHY and SORCE missions \citep{Skupin2005,Rottman2005}. SCIAMACHY on ENVISAT has been observing the solar spectrum daily since August 2002 and spectral data from SIM (Spectral Irradiance Monitor) on SORCE, are available from April 2004. However, absolute radiometric calibrations are not performed on ENVISAT data \citep{PagaranWeber2009} whereas they are for SIM. Therefore, SIM constitutes the longest, most reliable and continuous wide-band spectral dataset to date and now covers a period of over six years. Although the dataset does not encompass a full solar cycle, the period considered represents a decline of approximately a third in TSI and is sufficiently long to observe longer-term cycle-length changes, at least over several years, in spectral irradiance and to make model comparisons. \cite{HarderFontenla2009} published results from SIM between April 2004 and February 2008 showing the spectral trends so far observed. They found that the magnitude of variation in the UV was much larger than previously estimated and that the visible region varied in opposition to TSI, which leads to interesting and unexpected atmospheric feedback when these spectral forcings are considered in climate models \citep{HaighWinnning2010}.

Here we utilise the Spectral and Total Irradiance REconstruction model, or SATIRE \citep{FliggeSolanki2000,KrivovaSolanki2003}, to investigate the difference between modeled and observed solar irradiance on scales of days to years. A previous comparison of SIM and TIM data was made by \cite{UnruhKrivova2008} over three solar rotations. Over this short period the model agrees well with observation especially between 400 and 1300 nm. Here we extend this approach to a period of 73 rotations between 21 April 2004 and 1 November 2009 during the declining phase of cycle 23 to solar minimum. Over this longer period variability due to the solar-cycle can be compared.

In the next section we first describe the model setup used to make a comparison with SORCE data; the data are described in section 3 along with the analysis and processing performed. In section 4 we first compare the TSI reconstruction with SORCE/TIM and the PMOD TSI composite while in section 5 we turn to a spectral comparison between the model and SORCE/SIM, first in detail in three specific spectral regions and then for all wavelengths over 200-1630 nm. In section 6 we dicuss the results and present conclusions.

\section{Modeling Irradiance}
We model solar irradiance using the SATIRE model specifically tailored to satellite era data. It is denoted as SATIRE-S to distinguish it from other versions of the model from hereon in; cf. \cite{KrivovaSolanki2011}. A detailed overview can be found in \cite{KrivovaSolanki2003} but here follows a brief description.

\begin{figure}
        \resizebox{\hsize}{!}{\includegraphics[angle=0]{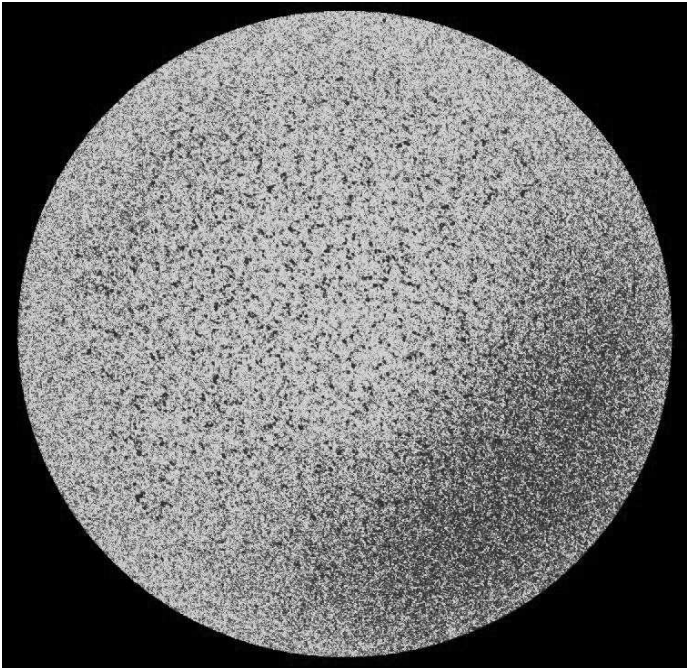}}
        \caption[]{
	A 5-min magnetogram showing $|$B$|$ taken by MDI on SoHO on 23 December 2008. Darker regions have a higher flux. Note the dark-area in the south-west quadrant, a bias present in all MDI magnetograms. 
        }
\label{fig:magnetogram}
\end{figure}

\begin{figure}
        \resizebox{\hsize}{!}{\includegraphics[angle=0]{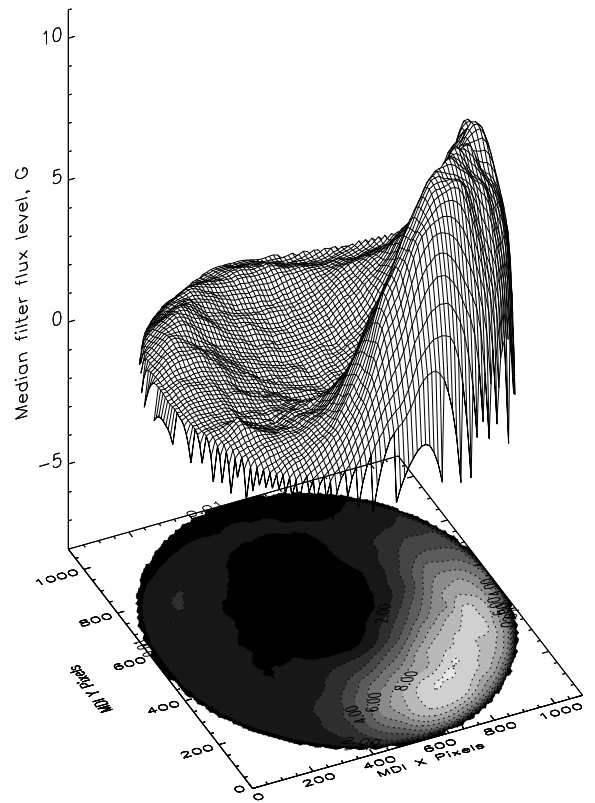}}
        \caption[]{
	The median filter used to remove the bias from magnetograms as seen in Fig.~\ref{fig:magnetogram}. The filter was formed by taking the mean of 100 images during the inactive period of 2008 and 2009 having applied a 41x41 pixel median window to each one. A scaled version of this filter is then used to subtract the bias from all magnetograms.
	        }
\label{fig:quad_filter_plot}
\end{figure}

SATIRE-S assumes that all variation in solar irradiance is the result of changes in the distribution of magnetic features on the solar surface. Any other changes to the Sun are not modelled, including variation in the intrinsic properties of the model components. If available, SoHO MDI full disk magnetograms and continuum intensity images\footnote{http://soi.stanford.edu/} \citep{ScherrerBogart1995} are used to identify four surface components which are each then assigned with an emergent intensity spectrum. Umbra and penumbra are identified in continuum images and faculae are identified in magnetograms where a significant magnetic signal is present, but no umbral or penumbral pixels are found in the continuum image. All remaining pixels are considered to be quiet Sun. The facular contrast is dependent on and proportional to the field strength of the magnetogram pixel up to a saturation point, $B_{\mathrm{sat}}$, beyond which the contrast is held constant. $B_{\mathrm{sat}}$ is the only free parameter in the model and is determined here by achieving a unity regression gradient with the comparison data\footnote{Regression slopes are calculated using the FITEXY routine from \cite{NumRecC}}. Both observation and theory suggest that increasing amounts of concentrated magnetic flux provide diminishing returns on contrast increases and $B_{\mathrm{sat}}$ takes this into account \citep{SolankiStenflo1985, FliggeSolanki2000, OrtizSolanki2002, Voegler2004}.

Previous reconstructions using MDI have used level 1.5 five-minute magnetograms \citep{KrivovaSolanki2003}. Here we use recalibrated daily level 1.8 averaged five one-minute magnetograms with a resulting 1$\sigma$ noise level of $\sim$13 G. We only consider signal above 3$\sigma$ of the noise.

The noise in MDI magnetograms is not uniform across the disk as can be seen in Fig.~\ref{fig:magnetogram}. One quadrant appears to have a consistently higher noise level. According to the Solar Oscillations Investigation website this is attributed to Doppler signal leakage\footnote{http://soi.stanford.edu/magnetic/Lev1.8/}. We attempt to correct this bias using a median-filter derived from 100 sunspot-free dates in 2008 and 2009. A similar approach was considered for older magnetograms in \cite{OrtizSolanki2002}. As can be seen in the resulting filter shown in Fig.~\ref{fig:quad_filter_plot} there is a maximum bias of 10 G in the magnetograms. Application of a scaled version of this filter results in a partial removal of the quadrant bias. There is an annual variation in the resulting lightcurve due to changes in the area of the Sun in that quadrant, but this has a negligable effect: a comparison between filtered and non-filtered TSI reconstructions showed little difference once the free parameter is determined (see below).

In the past, SATIRE \citep[e.g.,][]{WenzlerSolanki2004, KrivovaSolanki2011} considered facular flux in 5 G bins up to 1200 G, with all higher flux binned in this top level. We have found that a majority of the highest flux pixels have a contrast less than the quiet sun (though not dark enough to be considered as penumbra). Evidence suggests that the contrast of faculae is dependent on both the level of magnetic flux and on their disk location. At high flux levels exceeding 400G it has been found that for a fixed limb angle close to the disk centre contrast decreases and becomes negative \citep[][ P. Kobel, S. K. Solanki \& J.M. Borrero, in preparation]{TopkaTarbell1992, Voegler2004}. As faculae rotate towards the limb the contrast initially increases before declining, although the rate of change of contrast is dependent on the magnetic flux level \citep{TopkaTarbell1997,OrtizSolanki2002}. In particular, pixels with flux above a conservative estimate of 800 G do not display enhanced brightness. Until a physical flux, limb-angle and wavelength-dependent contrast \citep{AframUnruh2010b} can be employed these high flux pixels above 800 G are set to be quiet sun pixels and ignored. Also, some of the misidentification is a result of the line-of-sight correction for magnetic flux. The effect of this correction is most extreme at the limb and, as in all earlier versions, all pixels with a limb angle $\mu$ $< 0.1$ are ignored. This region accounts for only $\sim$1\% of the disk.

Once the filling factors of all pixels have been determined, intensities derived from model atmospheres are applied to each pixel type \citep{UnruhSolanki1999}, taking into account distance from the limb. The model atmospheres for quiet Sun, penumbral and umbral pixels use ATLAS9 \citep{Kucurz1993} assuming temperatures of 5777, 5400 and 4600 K respectively; the facular pixels use a modified FAL-P model atmosphere \citep{FontenlaAvrett1993}. The disk is then integrated over to produce a solar spectrum for that time and date. The resulting spectra span the range 200-160,000 nm and by repeating this for a series of dates, irradiance variations can be derived.

TSI data from the Total Irradiance Monitor (TIM) are available for comparison with SATIRE-S on 2141 days between 25 February 2003 and 1 November 2009, a period during which solar activity declined from intermediate to the depth of the recent minimum before starting to rise again. For the Spectral Irradiance Monitor (SIM), the comparison is considered from 20 April 2004 onward with data available for 1758 days of the 2021 day period, long enough to investigate variations on rotational and cycle-length timescales.

The free parameter, $B_{\mathrm{sat}}$, was determined to be 350 G by comparison with TIM. The regression of TIM with SATIRE-S is shown in Fig.~\ref{fig:tsi_regression} (black crosses) and is discussed in more detail in section~\ref{sec:sattim}. $B_{\mathrm{sat}}$ is higher than that obtained by \cite{KrivovaSolanki2003} due to the new calibration of MDI magnetograms. We fix $B_{\mathrm{sat}}$ to this value for all comparisons made in this paper.

\section{Irradiance Data}
\label{sec:observations}
The irradiance data used comes primarily from the SORCE satellite. Launched in January 2003, SORCE has four instruments which observe total and spectral irradiance variations \citep{Rottman2005}. Our analysis focuses on a comparison with two of these: total irradiance with TIM and spectral irradiance with SIM.

\subsection{SORCE/TIM}
\label{sec:tim}
The Total Irradiance Monitor (TIM) is a radiometer measuring the total solar irradiance with an absolute accuracy of 100 ppm and an instrumental noise level of $\sim$2 ppm. It has a stability of $<$10 ppm/yr \citep{KoppLawrence2005, KoppLawrence2005b}. Here Version 10 of the six-hourly cadence data is employed. We note that although the absolute value of TSI observed by TIM, of $\sim$1361 Wm$^{-2}$ is $\sim$4 to 5 Wm$^{-2}$ lower than other recent radiometers, the relative variation agrees to a high degree of accuracy. For our analysis the latter property is the important one, since SATIRE is designed to model variations in irradiance with a high a degree of accuracy and not the absolute value. TIM is used to fix the free parameter in the model, to test the accuracy of TSI variations from SATIRE-S and as a basis for considering the long-term plausibility of both wavelength-integrated SIM and SATIRE-S.

\subsection{SORCE/SIM}
\label{sec:sim}
The Spectral Irradiance Monitor, SIM, along with SCIAMACHY \citep{Skupin2005}, is one of the first instruments purpose-built to continually monitor solar UV, visible and IR radiation. The precision and measurement drift stability of SIM is at the 100 ppm level and it has an absolute accuracy of 2\% \citep{HarderThuillier2010}. We consider SIM data observed by three of the four photodiode detectors: the UV photodiode observes 200-308 nm, vis1 310-1000 nm and the IR 994-1655 nm, though we only use available data up to 1630 nm. This range covers $\sim$90\% of the spectral contributions to TSI. Relative accuracy is wavelength dependent: it is worst with 0.5\% at 310 nm, but above 500 nm it is better than 300 ppm \citep{HarderFontenla2005a}. Here we employ version 17 level 2 data with twice daily observations. For a space-based satellite, degradation of the instrument will occur on a long-term basis and SIM uses an onboard correction system that compares two spectrometers, SIM A and SIM B, to correct for long-term systematic degradation of the instrument (see appendix in \cite{HarderFontenla2009} for more detail).

\begin{figure}
        \resizebox{\hsize}{!}{\includegraphics[angle=90]{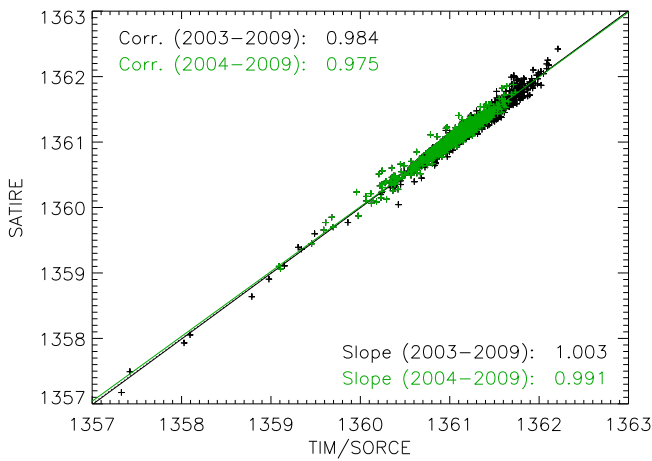}}
	\caption[]{Regression of TIM and SATIRE-S data over the period 28 February 2003 to 1 November 2009 (black) and for 21 April 2004 to 1 November 2009 (green). The single free parameter is varied until a regression slope close to 1.00 is found for the longer period. The dotted line representing a perfect fit is hidden behind the 2003-2009 black regression line.}
\label{fig:tsi_regression}
\end{figure}

\begin{figure*}
        \resizebox{\hsize}{!}{\includegraphics[angle=90]{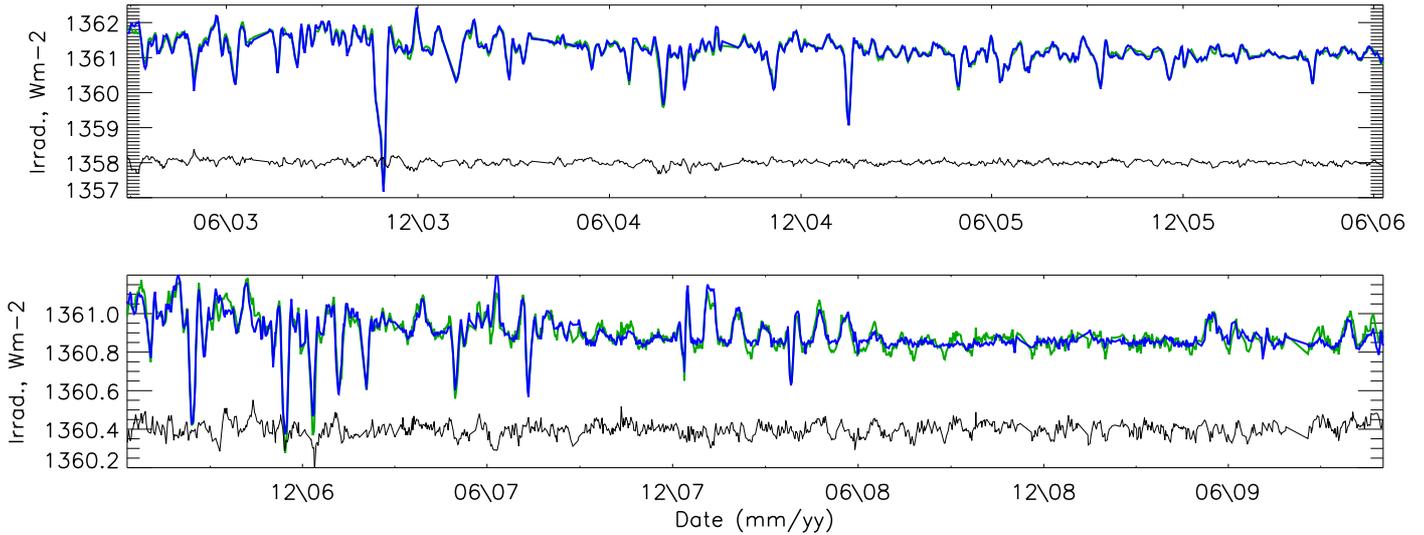}}
        \caption[]{
Irradiance lightcurves of TIM (green) and SATIRE-S integrated over 200-160,000 nm (blue) between 25 February 2003 and 1 November 2009. The upper plot shows the first half of the period and the irradiance scale covers 5.5 Wm$^{-2}$ while the lower half is on a reduced irradiance scale of 1.0 Wm$^{-2}$ to highlight the differences in the quieter minimum period of 2008 and 2009. The black line shows the difference between TIM and SATIRE-S TSI, shifted up by 1358 Wm$^{-2}$ in the upper frame and by 1360.4 Wm$^{-2}$ in the lower.}
\label{fig:tsi_lightcurves}
\end{figure*}

\subsubsection{Data preparation}
\label{sec:dataprep}
Before comparing the data to SATIRE-S we process it as follows. First, dates known in SIM to have instrument glitches are removed from the dataset. At some wavelengths severe glitches remain. Binomial smoothing \citep{MarchandMarmet1983} is used to produce an effective mean by smoothing the data and dates with flux values outside 7$\sigma$ of the instrument noise from the mean are identified. Dates with glitches in more than 200 wavelength elements and wavelengths with more than 300 days of glitches are identified and removed. Remaining individual elements are interpolated over from dates on either side. The process is then repeated at 5$\sigma$ to capture smaller but similarly erroneous outliers, more easily identified now the larger ones have been removed. These processes removes 0.6\% of the 3623 time series elements and 1.6\% of the 1829 SIM wavelength elements. Finally all outliers above or below 3$\sigma$ are interpolated over, so 7.0\% of the remaining dataset are interpolated over. At this stage no smoothing of the data has occured, other than interpolation over glitches.

SIM oversamples by 6 wavelength elements per resolution element, so a function simulating the resolution is applied to the spectrum. This reduces the number of spectral elements to 226. The same window is applied to SATIRE-S with the centre wavelength matching SIM. Binomial smoothing with a [1,2,1] filter is then performed on each remaining SIM time series to dampen random fluctuations.

Finally, the SATIRE-S and SIM data sets need to be put on the same time-grid. As the SATIRE-S dataset has a lower cadence it is used as a basis. SIM data are averaged over 24 hour periods and interpolated onto the SATIRE-S time-grid over the full seven-year period available. This, therefore, constitutes the procedure applied to SIM and SATIRE-S for comparison (for this paper it will be referred to as the `original' dataset to differentiate it from `detrended' and `smoothed' versions).

\section{Comparison of SATIRE-S and TIM Observations}
\label{sec:sattim}
Here we compare SATIRE-S integrated over all wavelengths (`SATIRE-S TSI') with TSI observations from TIM. The regression gradient of nearly 1.00 between SATIRE-S TSI and TIM produces a correlation coefficient, $r$, of 0.984 and so the model is able to explain 97\% of the variability (see Fig.~\ref{fig:tsi_regression}). Fig.~\ref{fig:tsi_lightcurves} shows the lightcurves of TIM (green) and the best fit SATIRE-S TSI (blue). The difference, as shown in black after shifting upwards for better visibility, is very small and has a standard deviation of just 0.06 Wm$^{-2}$. SATIRE-S recreates TSI well during active periods as shown in the upper frame, but during the minimum (lower frame, with a smaller scale), any low-level decaying weak magnetic flux of active regions remaining from the end of the current solar cycle are not picked up in five-minute magnetograms so that the small rotational variation is not well reproduced, e.g. from May 2008 to April 2009. Therefore, when the period considered is reduced to cover the timeframe of SIM from 2004 the correlation coefficient drops very slightly to 0.975, as does the regression slope, due to the removal of the strongest spots. Note that the level of difference between measured and reconstructed TSI does not change significantly with time, just the level of variability strongly decreases. This emphasises the importance of the long time spans that include periods of high (and low) activity to fix the free parameter. However, it is clear throughout that SATIRE-S recreates TSI very well.

As a further test, we repeat the same process with version d41\_62\_1003 of the PMOD composite \citep{Frohlich2000, Frohlich2006}. PMOD is a continuous composite of several satellite radiometers running over 32 years from 1978. PMOD data have been formatted to daily averages in contrast to the model which uses one set of data from a different time on each day. For the period considered, PMOD is exclusively composed of SoHO/VIRGO data, so that the comparison is effectively made with this instrument alone. Table~\ref{tab:tsi_compares_1} compares PMOD, TIM and SATIRE-S over the longer period 28 February 2003 to 1 November 2009 and the shorter period of 21 April 2004 to 1 November 2009 over which SIM is considered. The agreement between all three data sets is extremely good. The fact that the agreement between SATIRE-S and the observational data sets is almost as good as between TIM and PMOD implies that part of the remaining disagreement is due to inherent uncertainties in the observed data. Note that the poorer agreement with PMOD is in part due to not explicitly adjusting the free parameter for this data set. Achieving a slope close to 1.00 only requires a small change of $B_{\mathrm{sat}}$ to 340 G, with no significant change in correlation coefficient.

The value of $r^{2}$ estimates how well one dataset captures variability in the other. Since all three data sets over the longer period have similar levels of $r^{2}$ the success of the model at recreating TSI is emphasised. It should be noted that the higher correlation coefficients of all comparisons over the longer period (top, Table~\ref{tab:tsi_compares_1}) is the result of higher activity in 2003 providing a larger range of well-matched variability. There is only a small change in the results when the shorter period (bottom, Table~\ref{tab:tsi_compares_1}) is considered. This period is the same over which SIM will be compared with SATIRE-S and so the results here provide an upper limit and basis to which the wavelength-integrated irradiances can be compared. The comparison of SATIRE-S TSI and TIM over the shorter interval is marked in Fig.~\ref{fig:tsi_regression} by green crosses. Despite the lack of rotational variation identified between 2008 and 2009, SATIRE-S reproduces the TSI variation to an extremely high degree.

\begin{figure*}
        \resizebox{\hsize}{!}{\includegraphics[angle=90]{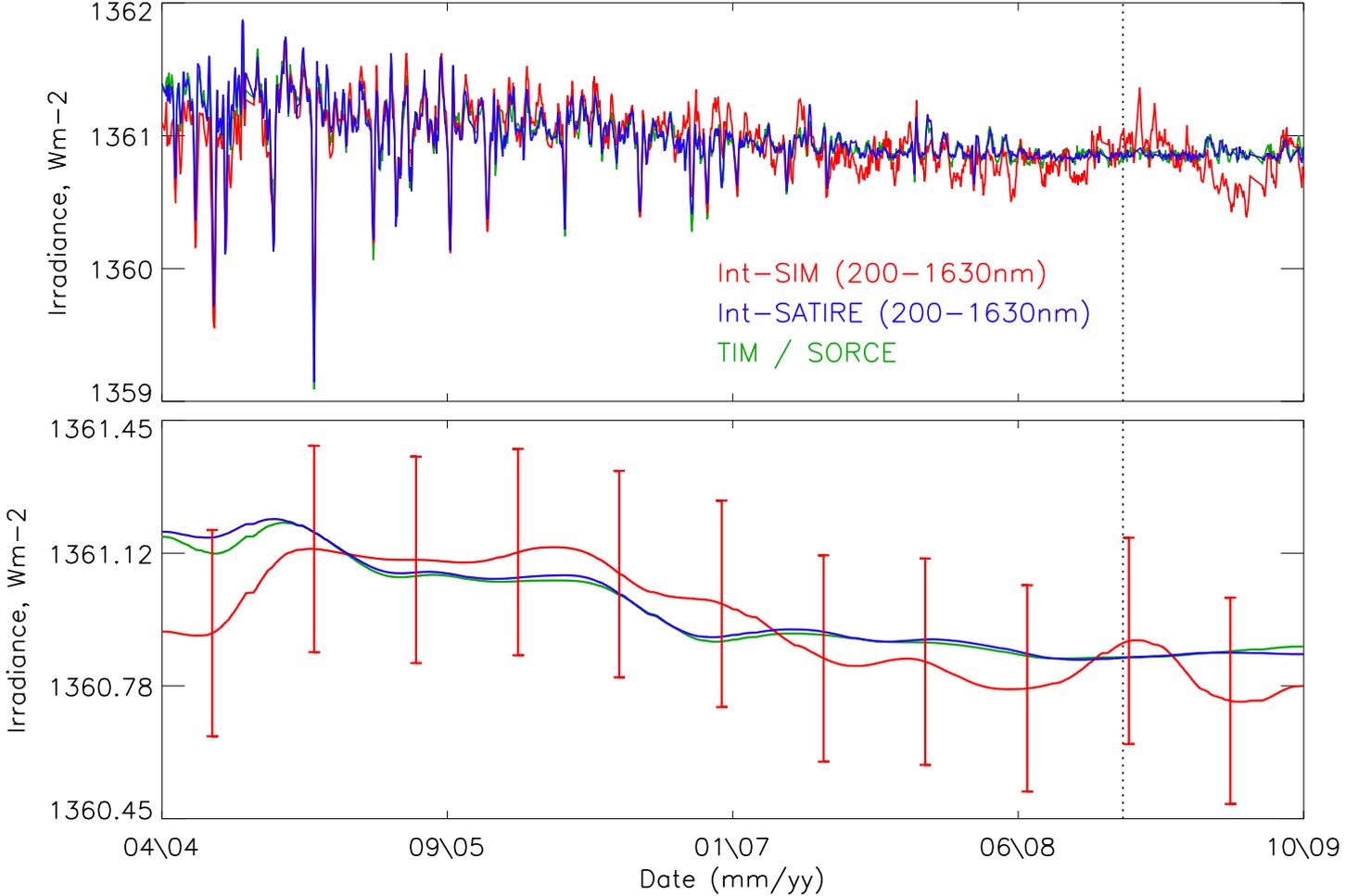}}
        \caption[]{Lighcurves of Int-SIM (red, 200-1630 nm), Int-SATIRE (blue, 200-1630 nm) and SORCE/TIM (green) for (top) original data and (bottom) smoothed or long-term data. Int-SIM and Int-SATIRE have a constant shift, given in Table~\ref{tab:tsi_compares}, added to bring them in line with TIM. Note that the y-axes are on different scales: the bottom panel displays the smoothed lightcurves from the top panel, with a scale of 3 Wm$^{-2}$, with a reduced irradiance scale of 1 Wm$^{-2}$. Error bars represent one standard deviation in the long-term stability of Int-SIM and are 0.259 Wm$^{-2}$ or 212 ppm. The vertical dotted line represents the time of solar minimum, December 2008.
        }
\label{fig:int_tsi_lcs}
\end{figure*}

\begin{figure*}
        \resizebox{\hsize}{!}{\includegraphics[angle=90]{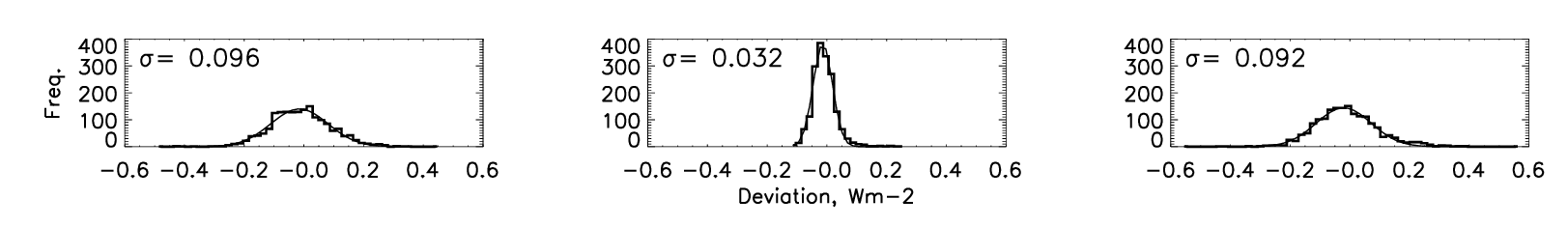}}
        \resizebox{\hsize}{!}{\includegraphics[angle=90]{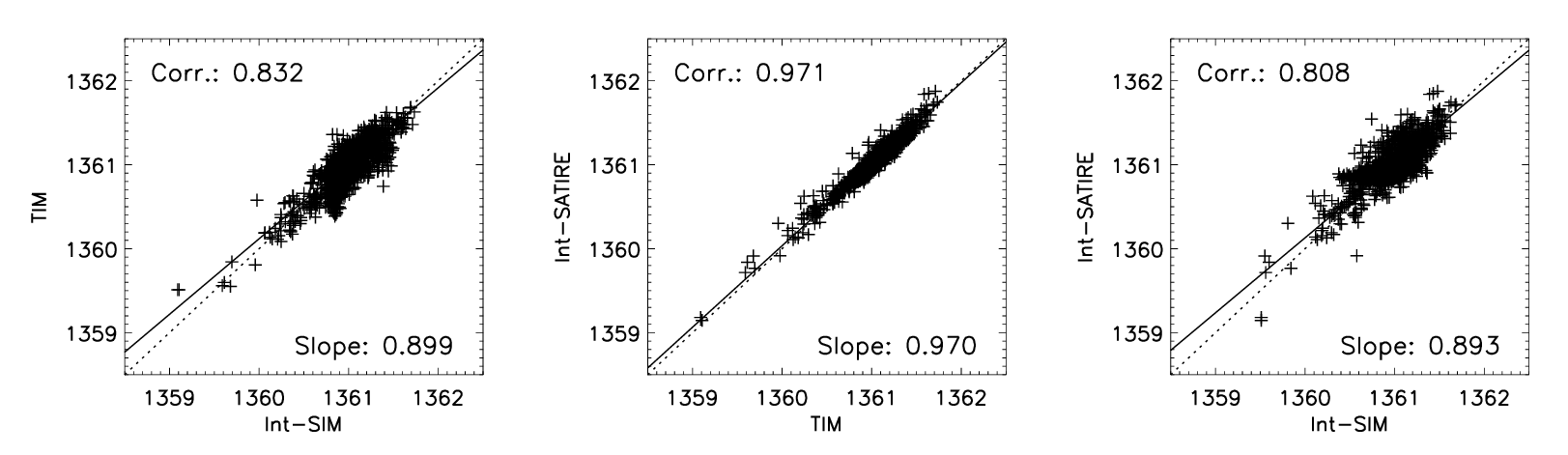}}
        \caption[]{Regression plots for left: Int-SIM vs. SORCE/TIM, centre: Int-SATIRE vs. SORCE/TIM and right: Int-SIM vs Int-SATIRE. Here Int-SIM and Int-SATIRE have had an additive constanst applied to bring their irradiance levels into line with TIM (see Table~\ref{tab:tsi_compares}). Axes in the regression plots all have units of Wm$^{-2}$. Above the regression plots are histograms of the spread about the regression lines. A Gaussian is fit to the histograms and the standard deviation of the Gaussian is presented in the top left.
        }
\label{fig:int_tsi_regs}
\end{figure*}

\begin{table}
\caption[]{\bf{Comparison of TSI between PMOD, SORCE/TIM and SATIRE-S.}}
\begin{center}
\begin{tabular}{llcr}
\hline
\multicolumn{4}{c}{28 February 2003 to 1 November 2009} \\
\hline 
Set 1 &   Set 2  &   $r   [r^2]$   &  Slope \\ \hline 
PMOD  &   TIM    &   0.987 [0.975] &  1.023 \\
TIM   &   SATIRE-S &   0.984 [0.969] &  1.003 \\
PMOD  &   SATIRE-S &   0.976 [0.953] &  0.980 \\ \hline 
\hline 
\multicolumn{4}{c}{21 April 2004 to 1 November 2009} \\
\hline 
Set 1 &   Set 2  &   $r   [r^2]$   &  Slope \\ \hline 
PMOD  &   TIM    &   0.977 [0.954] &  1.050 \\
TIM   &   SATIRE-S &   0.975 [0.951] &  0.991 \\
PMOD  &   SATIRE-S &   0.962 [0.926] &  0.944 \\ \hline 
\end{tabular}
\end{center}
\label{tab:tsi_compares_1}
\end{table}

\section{Comparison of the SATIRE-S model with SIM Observations}

In the following comparative analysis of SIM and SATIRE-S we present lightcurves of the data over various wavelength ranges. These are produced by integrating over the desired wavelengths and are broken down into `original', `smoothed' and `detrended' lightcurves. The smoothed lightcurves are derived by repeatedly binomially-smoothing with a [1,2,1] filter until short period fluctuations are eliminated. Smoothing does not occur over data gaps of greater than one day and these breaks are highlighted by small, horizontal discontinuities in the smoothed curve. The detrended time series shows only relative short-term variations and is obtained by dividing the original lightcurve by the smoothed one.

We consider that focus should be given to the correlations for discussion of the short-term; regressions should be considered for discussion of the long-term variability. To highlight this, the example of a high correlation coefficient but poorly matching long-term trend in the UV 200-300 nm region explains nothing about how well they agree. The magnitude of the long-term change in this example is very much in disagreement and only the regression picks this up (see section~\ref{sec:uv201}). A good regression for any detrended spectral bandwidth should be expected if the long-term trend is sufficiently removed and short-term variability is reasonably well recreated.

\subsection{Integrated Total Solar Irradiance}
\label{sec:int-tsi}
\begin{table}
\caption[]{\bf{Comparison of SORCE/TIM, Int-SIM and Int-SATIRE.}}
\begin{center}
\begin{tabular}{lllcr}
\hline
\multicolumn{5}{l}{Normalisation by Multiplication} \\
\hline 
Set 1      & Set 2      & Multip. factor & $r [r^2]$   &  Slope  \\ 
\hline 
Int-SIM    & TIM & 1.115 & 0.832 [0.692] &  0.774  \\
Int-SATIRE & TIM & 1.128 & 0.971 [0.944] &  0.909  \\
Int-SIM    & Int-SATIRE & - & 0.808 [0.653] &  0.904  \\ 
\hline 
\hline 
\multicolumn{5}{l}{Shift by Addition} \\
\hline 
Set 1 &   Set 2  & Add., Wm$^-2$ & $r [r^2]$   &  Slope  \\
\hline 
Int-SIM & TIM & 140.8 &  0.832 [0.692] &  0.899  \\
Int-SATIRE & TIM & 154.5 &  0.971 [0.944] & 1.033 \\
Int-SIM & Int-SATIRE & - &  0.808 [0.653] &  0.893  \\
\hline 
\end{tabular}
\end{center}
\label{tab:tsi_compares}
\end{table}

Before breaking the solar spectrum down into small wavelength bands, we compare `pseudo-TSI' measurements of SIM and SATIRE-S with TIM. When the SIM data are integrated over the 200-1630 nm range (`Int-SIM') they account for $\sim$90\% of TSI with a flux level of $\sim$1220 Wm$^{-2}$. This is in good agreement with SATIRE-S integrated over this region (`Int-SATIRE') at 1206 Wm$^{-2}$. The remaining $\sim$140 Wm$^{-2}$ is mainly emitted in the IR.

To compare Int-SIM and Int-SATIRE with TIM, these two datasets need to be multiplied (normalised) to TIM or shifted by a constant. Normalisation should provide an upper limit of any additional variability atrributed to TSI from wavelengths outside the 200-1630 nm integrated region while addition by a constant will assume no overall variability. In reality there is likely to be some additional variability and offsetting by different wavelengths. In Table~\ref{tab:tsi_compares} the effect of normalisation by multiplication or addition of a constant is shown with values derived from the mean of a one year period centred on December 2008. The results indicate that the addition of a constant provides a significantly better result, i.e. a slope closer to 1.00 and therefore a better approximation for the missing variability. This is supported by the evidence from the SIM electrical substitution radiometer (ESR) instrument, covering 1630-2400 nm, which shows no long-term trend over the period to solar minimum ($\sim$7x10$^{-4}$ Wm$^{-2}$) and is in line with the ever decreasing temperature sensitivity of the Planck function with increasing wavelength \citep{SolankiUnruh1998}. Since the addition of a constant better represents the overall missing variability than normalisation, comparisons hereafter are made considering an addition of a constant to align flux levels with TIM.

In Table~\ref{tab:int-tsi-corr} we present correlation coefficients and regression slopes for the three datasets separately for the original, detrended and smoothed data. Along with the lightcurve plots of Fig.~\ref{fig:tsi_lightcurves}, these provide an additional comparison of short- and long-term behaviour. Note that the correlations of TIM with Int-SATIRE differ from the SATIRE-S TSI results of section~\ref{sec:sattim} and Table~\ref{tab:tsi_compares_1} due to the restricted wavelength coverage of 200-1630nm. Furthermore, SATIRE-S data are now processed according to section~\ref{sec:dataprep} and the period considered is restricted to 21 April 2004 to 1 November 2009.

\begin{table}
\caption[]{\bf{Comparison between Int-SIM, Int-SATIRE and TIM for original, detrended and smoothed data.}}
\begin{center}
\begin{tabular}{llll}
\hline
\multicolumn{4}{l}{Original} \\ 
\hline 
Set 1        &   Set 2        &   $r [r^2]$   &   Slope  \\ 
\hline 
Int-SIM      &   TIM          &   0.832 [0.692] &   0.899		\\
TIM          &   Int-SATIRE   &   0.971 [0.944] &   0.970	\\
Int-SIM      &   Int-SATIRE   &   0.808 [0.653] &   0.893		\\ 
\hline 
\multicolumn{4}{l}{Detrended / Short-term} \\  
\hline 
Set 1        &   Set 2        &   $r [r^2]$   &   Slope  \\ 
\hline 
Int-SIM      &   TIM          &   0.868 [0.754] &   0.962		\\
TIM          &   Int-SATIRE   &   0.966 [0.934] &   0.949 		\\
Int-SIM      &   Int-SATIRE   &   0.842 [0.708] &   0.928 		\\   
\hline
\multicolumn{4}{l}{Smoothed / Long-term} \\  
\hline 
Set 1        &   Set 2        &   $r [r^2]$   &   Slope  \\ 
\hline 
Int-SIM      &   TIM          &   0.778 [0.605] &   0.766 		\\
TIM          &   Int-SATIRE   &   0.997 [0.994] &   1.051 		\\
Int-SIM      &   Int-SATIRE   &   0.764 [0.584] &   0.848 		\\ 
\hline
\end{tabular}
\end{center}
\label{tab:int-tsi-corr}
\end{table}

\cite{UnruhKrivova2008} found that the short-term period May - July 2004 showed a high correlation of 0.97 ($r^2 = 0.94$) between Int-SIM and TIM indicating very good short-term agreement. This period displays very little long-term variability and the majority of the correlation is captured by the short-term rotational variation from active region passages. In Table~\ref{tab:int-tsi-corr} we find that the correlation between Int-SIM and TIM of the original data is reduced to 0.83 ($r^2 = 0.69$) over the six-year period. This is the result of the variability dropping below the instrument noise level during the recent extended minimum (see below).

Fig.~\ref{fig:int_tsi_lcs} shows both the original (top) and smoothed (bottom) Int-SIM (red) and Int-SATIRE (blue) shifted by a constant as in Table~\ref{tab:tsi_compares} to bring them in line with TIM (green). The y-axis of the smoothed plot is smaller than the original data to show the long-term changes more clearly. The error bars of 0.26 Wm$^{-2}$ (212 ppm) in the lower panel represent a one standard deviation in the estimated long-term stability of Int-SIM. This instrument stability error is derived by computing the difference between the two onboard spectrometers, SIM A and SIM B, at as similar wavelengths as is achievable. The level is comparable to the noise equivalent irradiance of the detector. There are approximately annual oscillations in the smoothed plot with large deviations in the early period and around the solar minimum (December 2008, dotted line). These result from inexact degradation corrections made more difficult by slowly varying shifts in the wavelength scale and two spacecraft anomalies in 2009. The degradation corrections used here are the best that could be achieved at the time of writing.

Table~\ref{tab:int-tsi-corr} shows that agreement between Int-SIM and TIM (original) is reasonable with a correlation coefficient of 0.83, although the regression slope of 0.90 for the original data set and 0.78 for the smoothed indicates long-term disagreement, as can also be seen from the smoothed datasets of Fig.~\ref{fig:int_tsi_lcs}. In contrast, Int-SATIRE shows very good agreement in both original and smoothed cases. It should be noted that disagreement between Int-SIM and TIM (and Int-SATIRE) is due to the disagreement clearly seen in the early period and annual oscillations, but the long-term trend remains within the error bars. Unfortunately, these are so large that the above statement is of limited value, as it is possible that Int-SIM actually increases over the time period considered instead of declining in line with TIM while remaining within the error bars.

\begin{table*}
\caption[]{
\bf{Comparison of SORCE/SIM with SATIRE-S for selected inetgrated bands.}
	  }
\begin{center}
\begin{tabular}{lcccccc}
\hline 
	& (1) & (2) & (3) & (4) & (5) & (6) \\ 
\hline 
	& $\overline{\rm{SIM}}$, Wm$^{-2}$ & $\Delta$SIM, Wm$^{-2}$ & $\overline{\rm{SATIRE-S}}$, Wm$^{-2}$ & $\Delta$SATIRE-S, Wm$^{-2}$ & $r [r^2]$ & Slope \\ 
\hline
\multicolumn{7}{l}{Integrated UV: 201-300 nm} \\ 
\hline 
Original  & 15.05 & 0.34 & 15.04 & 0.07 & 0.770 [0.593] & 0.214	\\
Detrended & - & - & - & - & 0.731 [0.534] & 1.089	\\
Smoothed  & - & 0.31 & - & 0.07 & 0.856 [0.732] & 0.205	\\ 
\hline 
\multicolumn{7}{l}{Integrated visible: 400-691 nm} \\ 
\hline 
Original  & 521.17 & -0.71 & 521.14 & 0.09 & 0.313 [0.098] & 0.204	\\
Detrended & - & - & - & - & 0.892 [0.795] & 0.950	\\
Smoothed  & - & -0.67 & - & 0.06 & -0.672 [0.452] & -0.096	\\ 
\hline
\multicolumn{7}{l}{Integrated IR: 972-1630 nm} \\
\hline 
Original  & 295.88 & -0.24 & 295.89 & 0.00 & 0.405 [0.164] & 0.148	\\
Detrended & - & - & - & - & 0.762 [0.581] & 0.776	\\
Smoothed  & - & -0.25 & - & 0.00 & 0.737 [0.543] & 0.041	\\ 
\hline
\multicolumn{7}{l}{Integration over 200-1630 nm} \\
\hline 
Original  & 1223.01 & 0.01 & 1222.91 & 0.37 & 0.806 [0.650] & 0.891	\\
Detrended & - & - & - & - & 0.842 [0.708] & 0.925	\\
Smoothed  & - & 0.02 & - & 0.32 & 0.756 [0.571] & 0.855	\\ 
\hline
\end{tabular}
\end{center}
\tablefoot{
\bf{Columns (1) and (3) show the mean flux level of SIM and SATIRE-S, respectively, in December 2008; (2) and (4) show the flux change in SIM and SATIRE-S; (5) is the correlation coefficient and its square; (6) is the regression slope.}
}
\label{tab:ssi-data}
\end{table*}

The lower plots of Fig.~\ref{fig:int_tsi_regs} show regression plots comparing Int-SIM, Int-SATIRE and TIM with the regression slope and correlation quoted within. The upper panels display histograms of the scatter about the regression line in the lower plots. The short-term variability dominates the correlation coefficient result of the original datasets and can be seen by comparing these with the detrended results of Table~\ref{tab:int-tsi-corr} since long-term changes in TSI are small compared to rotational variability. It also indicates that, even without IR longward of 1630 nm, Int-SATIRE still recreates the TSI accurately over the six year period and provides a good substitute. Short-term variability (Table~\ref{tab:int-tsi-corr}, middle) of Int-SIM with TIM gives a good result. As can be seen from the upper panels of Fig.~\ref{fig:int_tsi_regs}, on a short timescale Int-SIM tends to overestimate the short-term variations in TSI, in particular in the later phases of the cycle.

Overplotted on each histogram in Fig.~\ref{fig:int_tsi_regs} is a Gaussian fit with the standard deviation quoted in the upper-right of the plot. All three plots show a roughly normal distribution. The scatter is a factor of three smaller for Int-SATIRE and TIM, adding further weight to the success of the model in reproducing TSI, though there is a slight skew to the right owing to some facular overestimation in the model. The spread for the other two comparisons here is dominated by the internal variability of Int-SIM which is much larger than TIM or Int-SATIRE and hence the distributions are much more spread out. It is clear from Fig.~\ref{fig:int_tsi_lcs} that the quiet low S/N period during solar minimum may be the cause of the degraded correlation between Int-SIM and TIM, while the lack of signal in Int-SATIRE during this period further decreases agreement between Int-SIM and Int-SATIRE conveyed by the lower correlation of 0.84. Indeed, increased correlations are found when the period is restricted to more active times, e.g. ending in February 2007.

Of most interest are the smoothed results. The long-term variation of some spectral regions were unknown before the launch of SIM. The correlation of smoothed Int-SATIRE with smoothed TIM of 0.997 again highlights the success of the model. We find that Int-SIM correlates less well with both TIM and Int-SATIRE. As seen in the long-term trend of the bottom panel of Fig.~\ref{fig:int_tsi_lcs} this is at least partly caused by the oscillation visible in 2008 and 2009 out-of-phase with TIM and Int-SATIRE. 

There is also a clear difference between Int-SIM and TIM of $\sim$0.3 Wm$^{-2}$ during April/May 2004. Although the two original datasets differ by only $\sim$170 ppm, this difference means that while TIM sees a decrease in irradiance of $\sim$0.4 Wm$^{-2}$ between April 2004 and the solar minimum\footnote{The value by which the datasets are shifted is taken from an average over a year centred at December 2008. This differs from the value in Table~\ref{tab:ssi-data}, but is done to smooth out the large bump during that period and make a fairer comparison.}, Int-SIM shows a decrease of only $\sim$0.1 Wm$^{-2}$. By mid-2005 all three datasets have reached a similar level of irradiance variation and SIM follows the TSI trend within errors. The main part of the difference between TIM and Int-SIM occurs during the period of largest change in all spectral regions, but it should be noted that the difference is just within one standard deviation of the long-term stability. We will consider the early difference further in the following sections while discussing the spectral variation and trends.

\subsection{Spectral Solar Irradiance}
We now consider specific spectral regions. Some of the spectral bands highlighted in \cite{HarderFontenla2009} are examined before expanding our analysis to narrower regions over the whole 200-1630 nm spectrum. In Figs~\ref{fig:ssi-uv},~\ref{fig:ssi-vis} and ~\ref{fig:ssi-ir} the original, detrended and smoothed lightcurves are shown for the spectral regions 201-300 nm (UV), 400-691 nm (visible) and 972-1630 nm (IR), respectively. These are again complemented by correlation coefficients, regression slopes and a measure of the change in irradiance between the average flux in the first month of the period and during the solar minimum month, {\bf{with negative values representing increases over time}}, displayed in Table~\ref{tab:ssi-data}. In the lightcurve plots SATIRE-S spectra (blue) are normalised to SIM (red) using a one-year period centred on December 2008 as the reference date (dotted line in plots). Error bars represent the estimated long-term stability of the instrument over the integrated regions and are derived in the same way as for Int-SIM (see section~\ref{sec:int-tsi}).

\subsubsection{UV, 201-300 nm}
\label{sec:uv201}
\begin{figure*}
        \resizebox{\hsize}{!}{\includegraphics[angle=90]{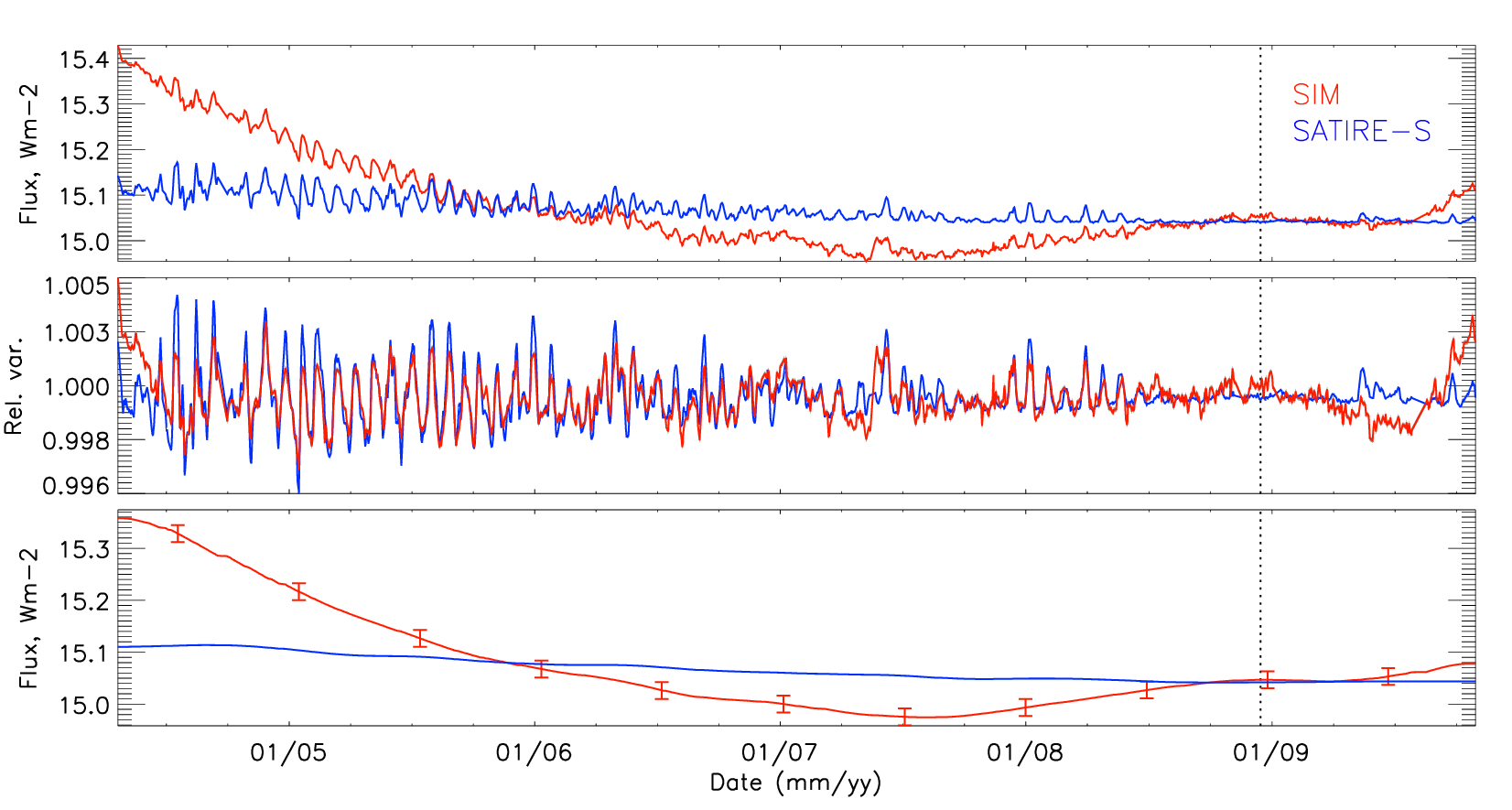}}
  			\caption[]{Lightcurves of the integrated UV region 201-300 nm for SIM (red) and SATIRE-S (blue). The top panel shows the original data, with SATIRE-S normalised by multiplication to SIM, while the middle and bottom panels show the detrended (short-term variations) and smoothed (long-term trends) curves.  One standard deviation error bars for the long-term stability of the instrument over these integrated wavelengths is 0.016 Wm$^{-2}$. The dotted line represents the solar minimum at December 2008 as in Fig.~\ref{fig:int_tsi_lcs}.}
  	\label{fig:ssi-uv}
        \resizebox{\hsize}{!}{\includegraphics[angle=90]{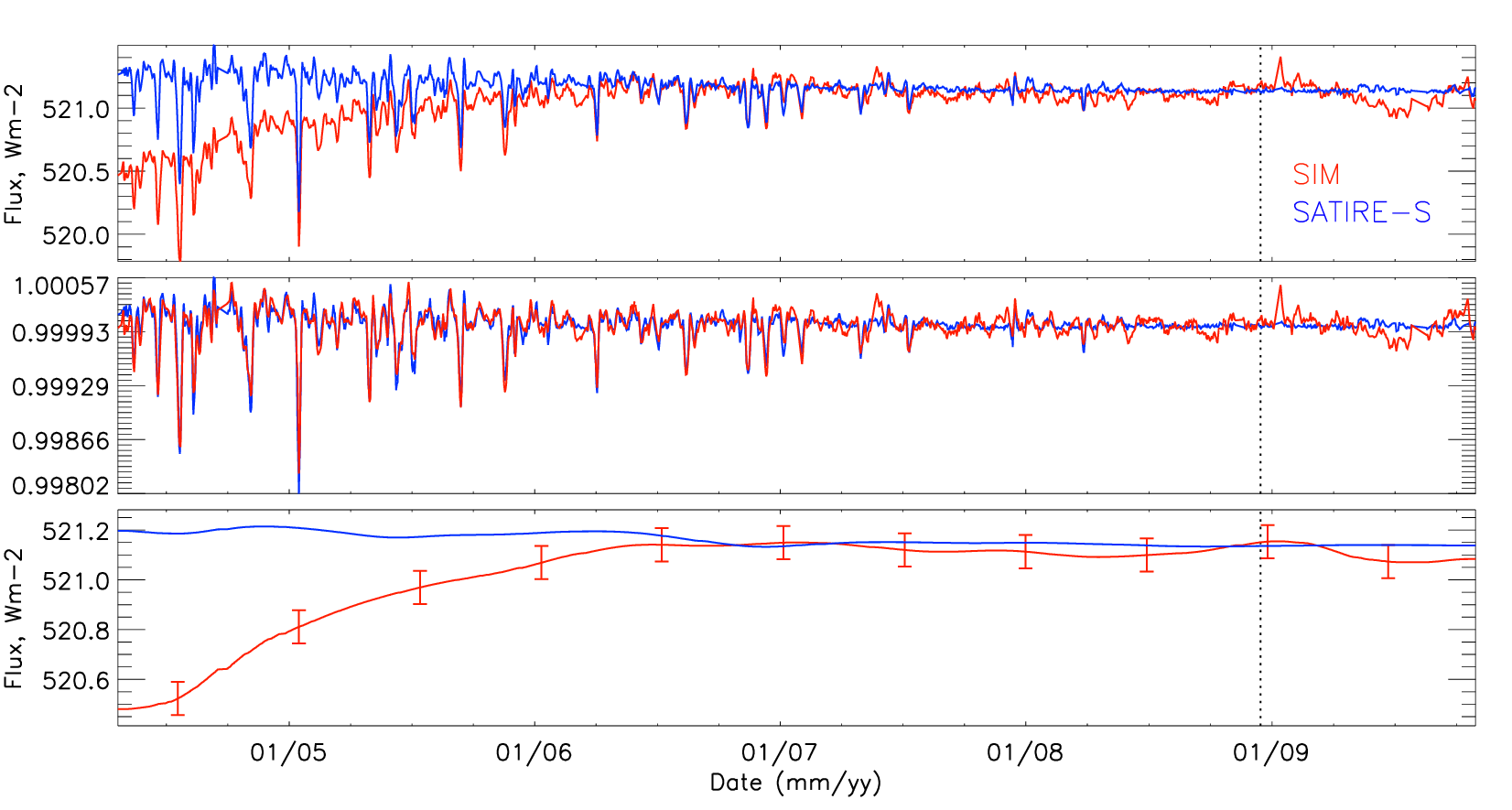}}
				\caption[]{As Fig.~\ref{fig:ssi-uv} but for the integrated visible region of 400-691 nm. Error bars are 0.067 Wm$^{-2}$.}
		\label{fig:ssi-vis}
\end{figure*}

\begin{figure*}
				\resizebox{\hsize}{!}{\includegraphics[angle=90]{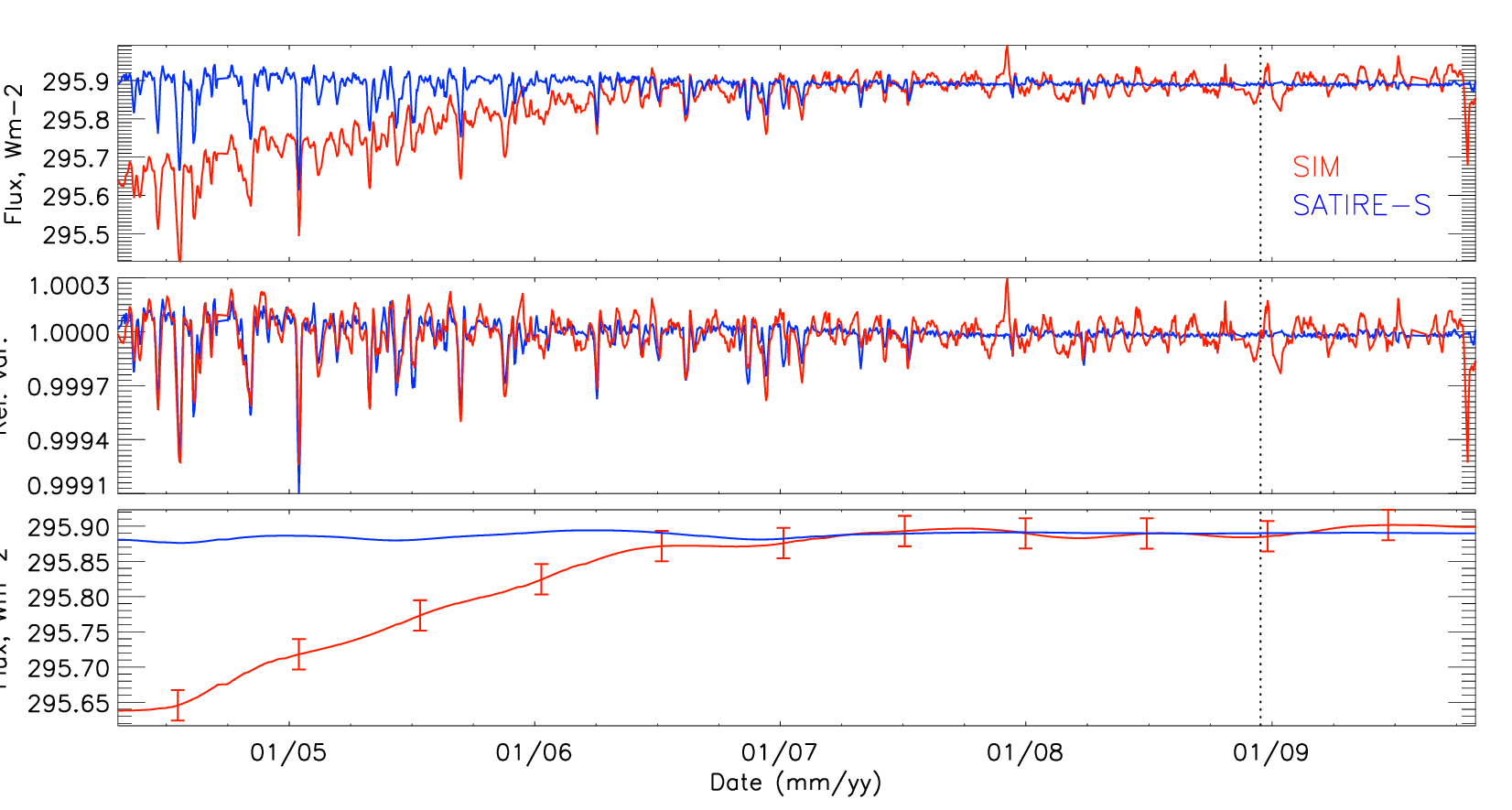}}
				\caption[]{As Fig.~\ref{fig:ssi-uv} but for the integrated IR region of 972-1630 nm. Error bars are 0.022 Wm$^{-2}$.}
		\label{fig:ssi-ir}
\end{figure*}

In the UV, the most striking difference between SIM and SATIRE-S is the difference in gradient seen in the original (top panel) and smoothed (bottom) lightcurves of Fig.~\ref{fig:ssi-uv}, in particular prior to the middle of 2006. The change in irradiance between 2004 and the reference date is 0.34 Wm$^{-2}$ for SIM, nearly five times larger than SATIRE-S's 0.07 Wm$^{-2}$ as presented in Table~\ref{tab:ssi-data}. This is also reflected in the regression slope of 0.21 in both the original and smoothed UV data. Note that the very high correlation in the smoothed data is probably the result of consistent agreement in the direction of change in the UV. However, although not clear in the smoothed UV lightcurve, a minimum is reached in SIM more than a full year prior to SATIRE-S and TSI measurements, in August 2007 and October 2008 respectively. The error bars establish that the disagreement between SIM and SATIRE-S exceeds the estimated long-term uncertainty of the instrument. It should be noted that given that the decline in TSI over this period is approximately a third of the maximum-to-minimum variation, this sets the cycle variation a significant factor above that observed by UARS/SUSIM and UARS/SOLSTICE \citep{FloydRottman2003, KrivovaSolanki2006}. Over a similar period in the previous cycle, August 1993 to May 1996 (solar minimum), the 200-300 nm region declined by 0.06 Wm$^{-2}$ from 14.51 to 14.45 Wm$^{-2}$ according to level 3BS V22 SUSIM data on UARS. The decline from maximum (November 1991) to minimum in that cycle was 0.11 Wm$^{-2}$ from 14.56 Wm$^{-2}$. Although the recent unusual minimum should be kept in mind, the change of TSI between recent minima has been very small, if it has changed at all \citep{Willson1997, DewitteCrommelynck2004, Frohlich2006, Frohlich2009b}. The reference date for the minimum is taken from sunspot number and it is interesting to note that according to SIM the UV minimum occured a year earlier, in 2007. 

It should be noted that by varying the free parameter, $B_{\mathrm{sat}}$, different gradients in the UV can also be achieved with SATIRE-S. A lower $B_{\mathrm{sat}}$ results in higher UV gradients (and also an incorrect TSI reconstruction), but even setting the value to a minimum the gradient seen in SIM cannot be reached and the resulting rotational variability becomes exceptionally large and unrealistic. In particular, SATIRE-S cannot match the huge drop in UV irradiance seen in SIM during the first two years. This early period also displays the largest change, larger than any other period considered, at all wavelengths and when integrated over the full range. This may be due to the fact that activity is declining to the minimum in 2008. Therefore, when spectral changes in SIM are discussed within this paper the vast majority of gradient and flux change occurs during this inital two-year period in the data set.

\begin{figure*}
        \resizebox{\hsize}{!}{\includegraphics[angle=90]{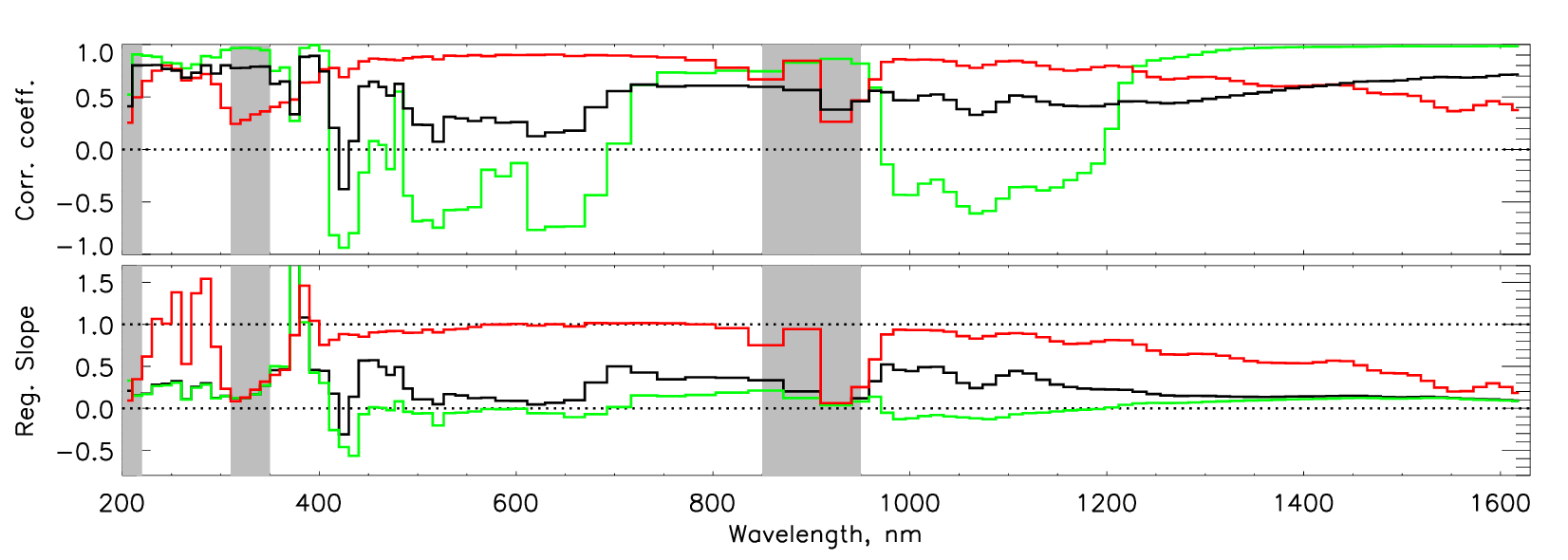}}
	\caption[]{Correlation coefficents (top) and regression slopes (bottom) between SIM and SATIRE-S as a function of wavelength for the spectrum from 200-1630 nm. For $\lambda <$ 450 nm wavelengths are compared in 10 nm bins; for $\lambda >$ 450 nm bins are resolution elements of SIM. There are three types of comparison made for each bin over the time period of 21 April 2004 to 1 November 2009: (black) original data, (red) detrended, short-term and (green) smoothed or long-term. Greyed-out regions highlight the detector edges where the signal is comparable to the instrument noise. Dotted lines are to aid the reader.}
\label{fig:ssi_cor_reg}
        \resizebox{\hsize}{!}{\includegraphics[angle=90]{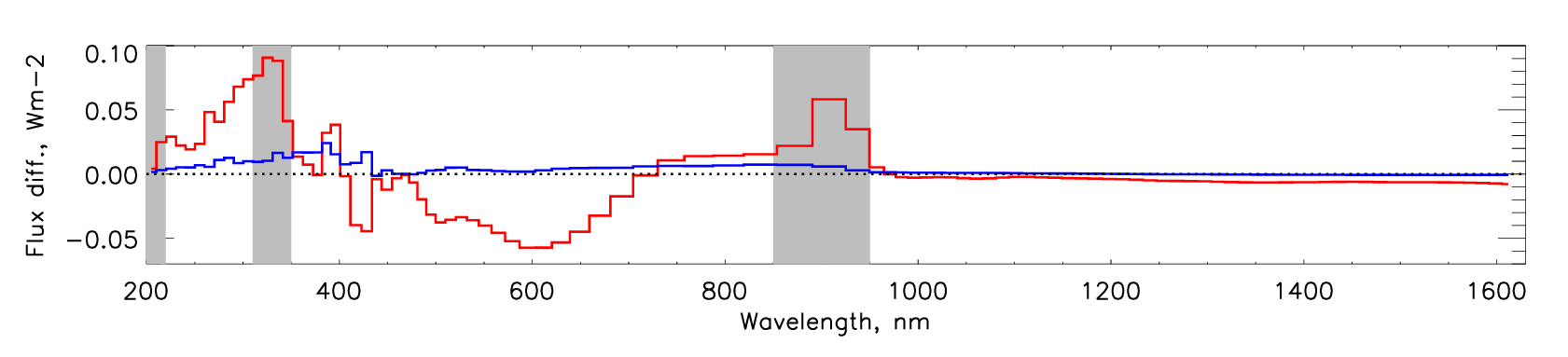}}
	\caption[]{Change in flux in the same bins as in Fig.~\ref{fig:ssi_cor_reg} of (red) SIM and (blue) SATIRE-S between mean level the first month of SIM's data and solar minimum in December 2008. Negative values imply an increase in irradiance over the period. The dotted line marks zero flux change and the greyed-out regions indicate the detector edges.}
\label{fig:ssi_flux_diff}
\end{figure*}

The rotational variation is already known to agree reasonably well when considered over a period of several months \citep{UnruhKrivova2008} and we also find that, over the six-year period considered here, rotational variation is well reproduced, though the magnitude is not always in agreement in particular in the IR at times close to activity minimum (Fig.~\ref{fig:ssi-uv}). The UV region reflects facular variation and we can see that SATIRE-S tends to overestimate that contribution, predominately during periods of higher activity. Between July 2008 and March 2009 the rotational variability observed by SIM is not recreated. This is a possible cause for the reduced correlation coefficients in this spectral region as compared to other spectral bands.

Facular overestimation occurs for two main reasons: the magnetic flux close to the limb is excessively amplified at locations where it is not vertical when applying a line-of-sight correction to the magnetograms, assuming all flux to be radial. Furthermore, the contrast applied to flux elements near the limb is also overestimated due to the use of one-dimensional model atmospheres \citep{SolankiUnruh1998}. \cite{UnruhKrivova2008} discuss the rotational variability in more depth and in comparison with the UARS/SUSIM instrument. They reported that below 240 nm and in some strong UV lines there is further disagreement since the intensties used to derive facular and spot contrasts are calculated on the assumption of LTE even where non-LTE would be more appropriate and so a difference is to be expected \citep{UnruhSolanki1999, KrivovaSolanki2005, HaberreiterKrivova2005, KrivovaSolanki2006}. It was also shown by \cite{ShapiroSchmutz2010}, using an updated version of the COSI radiative transfer code which performs calculations under non-LTE conditions \citep{HaberreiterSchmutz2008}, that LTE calculations can overestimate the facular contrast. Note that good agreement between SATIRE-S and TSI measured with SORCE/TIM on rotational timescales is found (see Fig.~\ref{fig:tsi_lightcurves}), although compensation by an underestimate at other wavelengths cannot be ruled out.

\subsubsection{Visible, 400-691 nm}
\label{sec:vis400}
Moving to the visible region at 400-691 nm (Fig.~\ref{fig:ssi-vis} and middle section of Table~\ref{tab:ssi-data}), a difference between observation and model is keenly seen in the long-term trend of original and smoothed lightcurves in the first two years. As noted in \cite{HarderFontenla2009}, there is an offset and steeply increasing trend in SIM for the early period in 2004 and 2005, in opposition to TSI, before flattening out towards the minimum. Again, the error bars in the SIM trend are much smaller than the trend itself. Over this period TIM TSI declines at a steady rate. SIM shows an increase in irradiance of 0.71 Wm$^{-2}$, almost an order of magnitude larger and in the opposite direction to SATIRE-S's 0.09 Wm$^{-2}$ decrease. The regression slope reflects this. Note that after mid-2006 the SIM trend and the model agree within the error bars.

In stark contrast to the long-term trends is the agreement seen for the detrended, rotational variability, indicated by the correlation coefficient of 0.89. This means that rotational variation and especially the spot and facular contribution is being well reproduced in the model. This can be seen in the detrended lightcurve where the magnitude of variation is clearly simulated well. The relative lack of signal detected in MDI images leads to a lack of rotational variability in SATIRE-S between April 2008 and April 2009, so that SATIRE-S underestimates the observed rotational variabiltiy over this time interval. It is the combination of clearly opposing early long-term trends and well matching short-term trends that results in the low correlation for the original data and the small regression slope seen in Table~\ref{tab:ssi-data}.

\subsubsection{IR, 972-1630 nm}
\label{sec:ir972}
A similar story for both short-term agreement and the longer-term trends as in the visible can be seen in the infra-red integrated between 972 and 1630 nm (Fig.~\ref{fig:ssi-ir}). It should be noted that the facular model employed by SATIRE-S results in faculae that become dark for $\lambda >  $1300 nm. As the region considered here captures wavelengths at which faculae are bright and dark, some cancellation will result in SATIRE-S over this region. In the longer-term, SIM observes an increase of 0.24 Wm$^{-2}$ and SATIRE-S registers no change. Inspection of the lightcurve during active times shows that the effect of spot passages are well reproduced, though there is uncertainty about whether facular effects are too. Apart from a couple of spot passages in 2008, from July 2007 to the end of the period, SATIRE-S is unable to reproduce the rotational variability seen with SIM, as could be seen to some extent in the visible, resulting in a lower correlation coefficient of 0.76. As in the UV, the low magnetic flux during this later period means that only a very small rotational amplitude is produced, too small to be seen on this scale.

\subsection{Comparison over all wavelength regions}
The approach taken in the previous section is now extended to narrow bands across the entire spectrum. Fig.~\ref{fig:ssi_cor_reg} shows the correlation and regression between SIM and SATIRE-S at all wavelengths from 200 to 1630 nm, in bins of 10 nm below 450 nm and at the instrument resolution of SIM above this wavelength for the period 21 April 2004 to 1 November 2009. 

In the upper and lower panels of Fig.~\ref{fig:ssi_cor_reg} we plot correlation coefficients and regression slopes as a function of wavelength, respectively. These are split into the original (black lines), detrended (red) and smoothed (green) datasets. Fig.~\ref{fig:ssi_cor_reg} is therefore a general overview of the spectral comparison of SIM and SATIRE-S. The grey regions highlight the detector edges where instrument noise is large and masks real signal.

In the UV we see that short-term variation below 230 nm is poorly recreated, identified by a correlation below 0.3. SATIRE-S assumes LTE across all wavelengths while this region is known to be dominated by non-LTE effects. This is also a region of where the instrument noise is comparable to variability in SIM. As wavelength increases, non-LTE effects become less important and LTE becomes a realistic assumption. In parallel the magnitude of solar variability increases relative to the instrument noise and is reflected in an improved short-term correlation. The short-term regression slopes highlight the regions of the UV where SATIRE-S overestimates the rotational variability during active periods, namely the 250-260 nm and 270-290 nm regions, again an effect of the LTE approximation, but also due to overestimation of the contrast near the limb. Up to wavelengths of 350 nm the magnitude of the gradient of the long-term trends do not agree, reflected by the regression slope close to zero.

For the vis1 photodiode above 350 nm short-term correlation coefficients become better than 0.90. Regression slopes remain close to zero in the long-term results, this being a result of opposing trends between SATIRE-S and SIM prior to 2006 and the smoothed late period agreement when both datasets show almost flat lightcurves (see Fig.~\ref{fig:ssi-vis} and section~\ref{sec:vis400}). In the detrended results the high correlation coefficient and the regression slope near to 1.0 across the entire visible region highlights how well the magnitude of the rotational variability is reproduced by SATIRE-S.

In the IR there is a gradual decline in correlation of the short-term variation, dropping from a correlation coefficient above 0.8 to below 0.4 near the edge of the detector at 1630 nm. We note that although spot passages are well identified in this spectral region, rotational variation due to faculae is not always. During periods of higher activity, and therefore higher magnetic flux, faculae in the IR can appear dark \citep{UnruhKrivova2008,FontenlaHarder2004}. But there appears to be a switchover to positive contrast in faculae of lower field strength, especially during lower activity. We see evidence for bright faculae in the IR from SIM during several rotations in 2008 that vary in phase with TSI and during which there were no spot passages. Over the IR domain there is a gradual change in the smoothed regression slope and correlation coefficient from negative to positive. This is because SATIRE-S assumes that faculae are all dark for $\lambda >$ 1300 nm, as mentioned in section~\ref{sec:ir972}, except near to the limb. Facular disk transits therefore result in reduced variability and ultimately produce a rotational variation in opposition to the flux measured in SIM in the later, quieter period. Integration over a wide IR spectral band, as in Fig.~\ref{fig:ssi-ir}, combines these regions of positive and negative contrast which cancel each other to produce a flat smoothed lightcurve.

In many regions throughout the entire spectrum, the effect of conflicting short- and long- term trends is highlighted by the divergence of both green (smoothed) and red (detrended) lines from the black (original) line. SATIRE-S cannot reproduce the magnitude or shape of the long-term trend in almost all spectral regions, though the accuracy of the observed long-term trends in SIM is still to be validated.

Fig.~\ref{fig:ssi_flux_diff} shows the contribution by each bin to the total change in Int-SATIRE and Int-SIM by taking the difference of the averaged 30-day period from 21 April 2004 to 21 May 2004 and the month of December 2008 at solar minimum. Fig.~\ref{fig:ssi_flux_diff} demonstrates that the offset that occurs between each spectral region is significantly less in SATIRE-S than with SIM. Although the integral over 200-1630 nm is not in agreement, the disagreement is within the error shown in Fig.~\ref{fig:int_tsi_lcs}. The missing 10\% of irradiance $>$1630nm does not vary enough, suggested by the ESR detector, to make up this shortfall. Looking at Fig.~\ref{fig:int_tsi_lcs}, the missing flux must therefore be from the disagreement in the 2004 and the large oscillation around December 2008. 

Unfortunately, periods considered in both \cite{HarderFontenla2009} and this paper are only during a declining period in TSI. Therefore, any reversal in these individual spectral regions, which might provide weight to plausibility of the spectral trends observed, cannot be considered for now.

\section{Discussion and conclusion}
\label{discussion}

In this paper we have made comparisons of total and spectral solar irradiance observations from the SORCE satellite, namely the TIM and SIM instruments \citep{KoppLawrence2005, HarderFontenla2005a}, with the SATIRE model \citep{FliggeSolanki2000, KrivovaSolanki2003}. We have broken down the broad spectrum from 200-1630 nm into small regions and compared short-term rotational variability and long-term trends over a period of 6 years from moderate solar activity in 2004 to the solar minimum in 2008 and 2009.

Our investigation shows that although SATIRE-S is highly successful at recreating total solar irradiance over both rotational and cycle length periods, there are large differences between observation and model in a number of spectral regions, specifically regarding the gradients of the long-term trends between 2004 and 2006.

Over the period between 2004 and 2009 considered in this paper SATIRE-S reproduces 95\% of TSI variation (97\% for 2003-2009), a level of agreement equal to that of different TSI instruments, while Int-SIM, i.e. SIM integrated from 200 to 1630 nm, accounts for 69\%. We find that Int-SIM shows excellent short-term agreement with TIM. The longer-term trends tend to agree reasonably well, though there is substantial diagreement in the first 12 months where there are rapid changes in all spectral regions and the largest variation in flux occurs.

On short timescales the SATIRE-S UV shows good agreement in capturing rotational variation, but overestimates the amplitude in some spectral regions during years of higher activity, e.g. at 250-260 nm and 270-290 nm. This results from overestimating the contrast for faculae close to the limb and the use of LTE approximation in a region where this is inadequate \citep{KrivovaSolanki2006}. It should be noted that the limb- and flux-dependent contrasts applied in SATIRE-S are based on one-dimensional models and are therefore not likely to model the physics fully. \cite{Voegler2004} has shown using MHD simulations that increasing the integrated magnetic flux within a facular region, while initially increasing in brightness, eventually sees a sharp decline to a negative contrast at very high magnetic flux levels. \cite{AframUnruh2010b} have also used results from similar simulations to find variation in the spectral contrast dependent on the level of magnetic flux in the region.

The rotational variability in the visible is the region best recreated by SATIRE-S, while short-term fluctuations are well matched with SIM in the IR during active periods. During the solar minimum in 2008-9 for $\lambda>$ 1300 nm a clear signal is present in the observational data while almost no variation is seen in SATIRE-S at any wavelengths. This reveals an issue when using MDI 5-minute integrated magnetograms. The signal of magnetic flux in these magnetograms during the 2008-9 period is extremely weak and the resulting modeled lightcurves are almost completely devoid of rotational variability whereas TIM shows small but clear variation.

In SATIRE-S, during active periods, the short-term over-enhancement of facular regions in the UV more than compensates for the negative contrast applied in the IR to these regions. MHD simulations have found that the IR contrast for low magnetic flux levels remains positive for $\lambda >$ 1300 nm in agreement with the detrended SIM results (N. Afram, private communication). This evidence highlights that not all of the wavelength-dependent contrasts applied in SATIRE-S to facular pixels detected in MDI magnetograms are correct and that a transition from positive to negative contrast in the IR for $\lambda >$ 1300 nm results in disagreement with the observations from SIM. Contrasts are dependent on wavelength, limb-angle and magnetic flux level \citep{FoukalMoran1994}. These variables change over the solar cycle and result in a different contribution to TSI. It is therefore important not only to improve the model rotational variability, but also to investigate the inter-cycle variation which is currently of great interest given the unusually quiet period the Sun is only now leaving. 

The long-term UV trends observed in SIM and computed by SATIRE-S disagree strongly. The magnitude of variation over the first two-year period, 2004-2006, in the UV 200-300 nm region is five times larger in SIM than SATIRE-S. There is some agreement of this trend in SIM with the SORCE/SOLSTICE\footnote{SORCE/SOLSTICE data were downloaded from http://lasp.colorado.edu/sorce/data/data\_product\_summary.htm on 17 September 2010.} \citep{McClintockRottman2005} UV instrument, though this depends on the period of overlap considered. The short- and long-term trends of UARS/SUSIM agree with SATIRE-S in 2004. But, between 2005 and 2007 SORCE/SOLSTICE disagrees with SATIRE-S while having a reasonable agreement with SIM. Post-2007, SIM starts to increase again while SORCE/SOLSTICE continues to decline. Therefore, no clear picture emerges as yet from a comparison with the SORCE/SOLSTICE data even though there is a decline of 0.27 Wm$^{-2}$ observed between April 2004 and December 2008, in reasonable agreement with SIM.

In the visible region, SIM observes an increase in irradiance as TSI decreases, large enough to offset the significant reduction in irradiance produced by the UV region. An increase in irradiance in this region as detailed in \cite{HarderFontenla2009} is in opposition to previously modeled spectral variation \citep{Lean1991, KrivovaSolanki2006} and to that used in global climate models \citep{HaighWinnning2010}. 

In the IR, except for the 750-900 nm region, long-term changes displayed by SIM are all in opposition to TSI and SATIRE-S for $\lambda <$ 1300 nm. As a result of most of the spectrum in SATIRE-S decreasing with TSI, the contribution from each region to the overall change is, in some regions, more than an order of magnitude less than the trends observed in SIM.

The model assumes that variations in irradiance are directly related to the evolution of surface magnetic flux. Given this, if the assumption is made that the long-term UV results from SIM are indeed correct then there are physical implications. Either the long-term change being observed is dominated by otherwise unknown non-magnetic processes or it is due to changes in small-scale flux which cannot be detected in 5-min MDI magnetograms. It would also imply a physical change in the Sun during the decline of this solar cycle as compared to the previous one, and even compared with the rising phase of cycle 23, if the results from UARS/SUSIM and UARS/SOLSTICE are correct. Recall that SATIRE-S reproduced the data from these instruments for the same value of the free parameter as needed for TSI. Hence, although the declining phase of cycle 22 and rising phase of cycle 23 UV variations are consistent with surface magnetism as a source, this appears no longer to be the case in the declining phase of cycle 23. Such a sudden change in the physical mechanism seems rather unlikely to us and implies that there might be a problem in the calibration or stability of either the UARS or the SORCE UV measurements. Indeed, a recent study \citep{KrivovaSolanki2011b} using 60-min MDI magnetograms in SATIRE-S shows a very similar long-term trend to the one shown here. In this study, weak magnetic flux is taken into consideration and better accounts for the effect of changes in surface magnetism. In this context it needs to be noted that the UARS instruments covered both the declining and rising solar cycle phases, whereas SORCE has so far mainly covered the declining phase. Also, the discrepancy of SIM to SATIRE-S is largely restricted to the first two years of SIM's operation. Many radiometric instruments studying the Sun in space have had problems maintaining the required level of stability in the first couple of years of observation.

As the period considered ends during the solar minimum, there is no possibility at this time of investigating the period of increasing activity. Any reversals in the SIM spectral regions on the return to solar maximum would provide support for the spectral trends observed. If SIM is able to continue observing for the next year or two then this may indeed be seen. SOLSPEC \citep{ThuillierFoujols2009}, on the International Space Station, is also monitoring the region observed by SIM, but it only became active during the solar minimum and results are still pending.

As a result of the investigation made within this paper, areas of improvement to the model have been highlighted. Primarily, the contrasts used in SATIRE-S must be addressed in a physical way to incorporate dependence not just on limb-angle but also on the level of magnetic flux and wavelength. Work is currently underway to address this using results from MHD simulations. It appears very unlikely, however, that such changes will be able to overcome the large difference in trend between SIM and SATIRE-S at most wavelengths. 

\section*{ACKNOWLEDGMENTS}
This work was supported by the \emph{Deut\-sche For\-schungs\-ge\-mein\-schaft, DFG\/} project number SO~711/1-3, by the NERC SolCli consortium grant and by the WCU grant No. R31-10016 funded by the Korean Ministry of Education, Science \& Technology. 

We also thank the International Space Science Institute (Bern) for giving us the opportunity to discuss this work with the international team on "Interpretation and modelling of SSI measurements".

In the use of version d41\_62\_1003 of the PMOD dataset from PMOD/WRC, Davos, Switzerland, we acknowledge unpublished data from the VIRGO Experiment on the cooperative ESA/NASA Mission SoHO.

We thank Borut Podlipnik for help putting together lists of MDI images that were used in this work and Hao Thai at the Stanford Solar Center for help preparing SOHO MDI data for download. SOHO Data supplied courtesy of the SOHO/MDI and SOHO/EIT consortia. SOHO is a project of international cooperation between ESA and NASA. We thank the referee for careful consideration of the paper and useful suggestions.

\bibliography{solarbib}
\bibliographystyle{aa}

\end{document}